%% file: eprint.tex
\documentclass[12pt]{article}
\usepackage{graphicx}
\usepackage{amsmath}
\usepackage{lineno}
\usepackage{caption}
\usepackage{hyperref}\hypersetup{hidelinks}
\newcommand{\Wp}{W^{\prime}}
\newcommand{\pT}{p_{\text{T}}}

\def\institute{Department of Physics and Astronomy\\
Michigan State University}
\def\support{\footnote{\copyright~2021 CERN for the benefit of the ATLAS Collaboration. Reproduction of this article or parts of it is allowed as specified in the CC-BY-4.0 license}}

\def\Title#1{\begin{center} {\Large #1 } \end{center}}
\def\Author#1{\begin{center}{ \sc #1} \end{center}}
\def\Address#1{\begin{center}{ \it #1} \end{center}}

\newenvironment{Abstract}{\begin{quotation}  }{\end{quotation}}
\newenvironment{Presented}{\begin{quotation} \begin{center} 
             PRESENTED AT\end{center}\bigskip 
      \begin{center}\begin{large}}{\end{large}\end{center} \end{quotation}}

\input econfmacros.tex

\begin{document}
\begin{titlepage}

\vfill
\Title{Search for $\Wp \to tb$ decays in the fully hadronic final state with the ATLAS experiment}
\vfill
	\Author{ Kuan-Yu Lin on behalf of the ATLAS Collaboration\support}
\Address{\institute}
\vfill
\begin{Abstract}
	A search for a new heavy boson $\Wp$ in proton-proton collisions at $\sqrt{s}$ = 13 TeV is presented. The search focuses on the decay of the $\Wp$ to a hadronic top quark and a bottom quark, using the full Run 2 dataset of the ATLAS detector. The hadronic decay of the top quark is identified using DNN-based boosted-object techniques. The dominant background is obtained by a data-driven method with small systematic uncertainties. The results are presented as upper limits on the production cross-section times decay branching ratio for the $\Wp$ boson with right-handed couplings that decays to a top quark and a bottom quark, for several $\Wp$ masses between 1.5 to 5 TeV.
\end{Abstract}
\vfill
\begin{Presented}
$14^\mathrm{th}$ International Workshop on Top Quark Physics\\
(videoconference), 13--17 September, 2021
\end{Presented}
\vfill
\end{titlepage}
\def\thefootnote{\fnsymbol{footnote}}
\setcounter{footnote}{0}

\section{Introduction}
A new heavy charged vector boson $\Wp$ is found in various BSM scenarios such as the Little Higgs models~\cite{Perelstein:2005ka} and extra-dimensional models~\cite{Cheng:2001vd}. In this analysis, the ATLAS collaboration~\cite{Collaboration_2008} conducts a resonance search for the $\Wp$ in the all-hadronic $tb$ final state ($t\bar{b}$ and $\bar{t}b$). The $tb$ channel is sensitive to $\Wp$ candidates with larger couplings with the third generation~\cite{Li:1992fi,Malkawi:1996fs,Muller:1996dj,Calabrese:2021lcz} or quarks~\cite{GEORGI1990541}. Hadronically decaying top-quarks are reconstructed as $R$ = 1 jets and identifiable via jet substructure techniques~\cite{Kogler:2018hem}.

\section{The jet definitions and the invariant mass \texorpdfstring{$m_{tb}$}{mtb}}
Using the Run-2 ATLAS data equivalent to 139 $fb^{-1}$, this search considers proton collision events with no leptons, at least one large-$R$ ($R$ = 1) jet with $\pT$ $>$ 500 GeV, and a small-$R$ jet ($R$ = 0.4) with $\pT$ $>$ 500 GeV in the opposite direction, as viewed on the plane perpendicular to the proton beams. The two jets corresponds to the top quark and bottom quark decaying from the $\Wp$. These two jets are summed, resulting in the invariant mass $m_{tb}$, which is expected to assume a smoothly falling distribution for the SM backgrounds. By identifying jets associated with the top and bottom quarks, a peaking structure around the $\Wp$ mass becomes significant.

\section{The identification of jets associated with a top quark or a bottom quark}
The energy pattern of the calorimeter cluster energy~\cite{ATLAS:2016krp} in a large-$R$ jet facilitates the top-quark tagging. Jet substructure variables such as the N-subjettiness~\cite{Thaler:2010tr,Thaler:2011gf} distinguish the energy deposits due to the three quarks of top-decay from a light parton's QCD radiation. Several substructure variables, the jet mass, and $\pT$ are combined via a Deep Neural Network (DNN) into a top-tagging score~\cite{ATLAS:2018wis}. A large-$R$ jet with $\pT$ $>$ 500 GeV and passing the 80\% Working Point (80\% WP, the threshold corresponding to an 80\% probability) is designated as a \textbf{top-candidate jet}. Events with more than one top-candidate jet are rejected for the suppression of the top-quark pair ($t\bar{t}$) background. In contrast, events without a top-candidate jet but with large-$R$ jets ($\pT$ $>$ 500 GeV) passing the DNN score above $e^{-7}$ are reserved for separate analysis regions. Such large-$R$ jets are called the \textbf{top-proxy jets}.

The highest $\pT$ small-$R$ jet among those with $\pT$ $>$ 500 GeV and $\Delta\phi$ $>$ 2 from the top-candidate jet is declared the \textbf{$b$-candidate jet}. Each top-proxy jet is also paired with a $b$-candidate jet. Despite bearing the letter $b$ in its name, the $b$-quark identification is employed later for region assignment. The ATLAS $b$-tagging algorithm called the DL1r~\cite{ATLAS:2019bwq} reconstructs the $b$-hadron decay vertices using charged particle tracks around the small-$R$ jet axis. The $b$-tagging requirement applied to the $b$-candidate jet is the 85\% WP. Since the top decay products include a $b$-quark, small-$R$ jets with $\pT$ $>$ 25 GeV inside the top-candidate (top-proxy) jet are checked by the same $b$-tagging requirement.

\section{Data-driven background estimation and region assignment}
\begin{figure}[tph]
  \centering
  \includegraphics[width=0.8\textwidth]{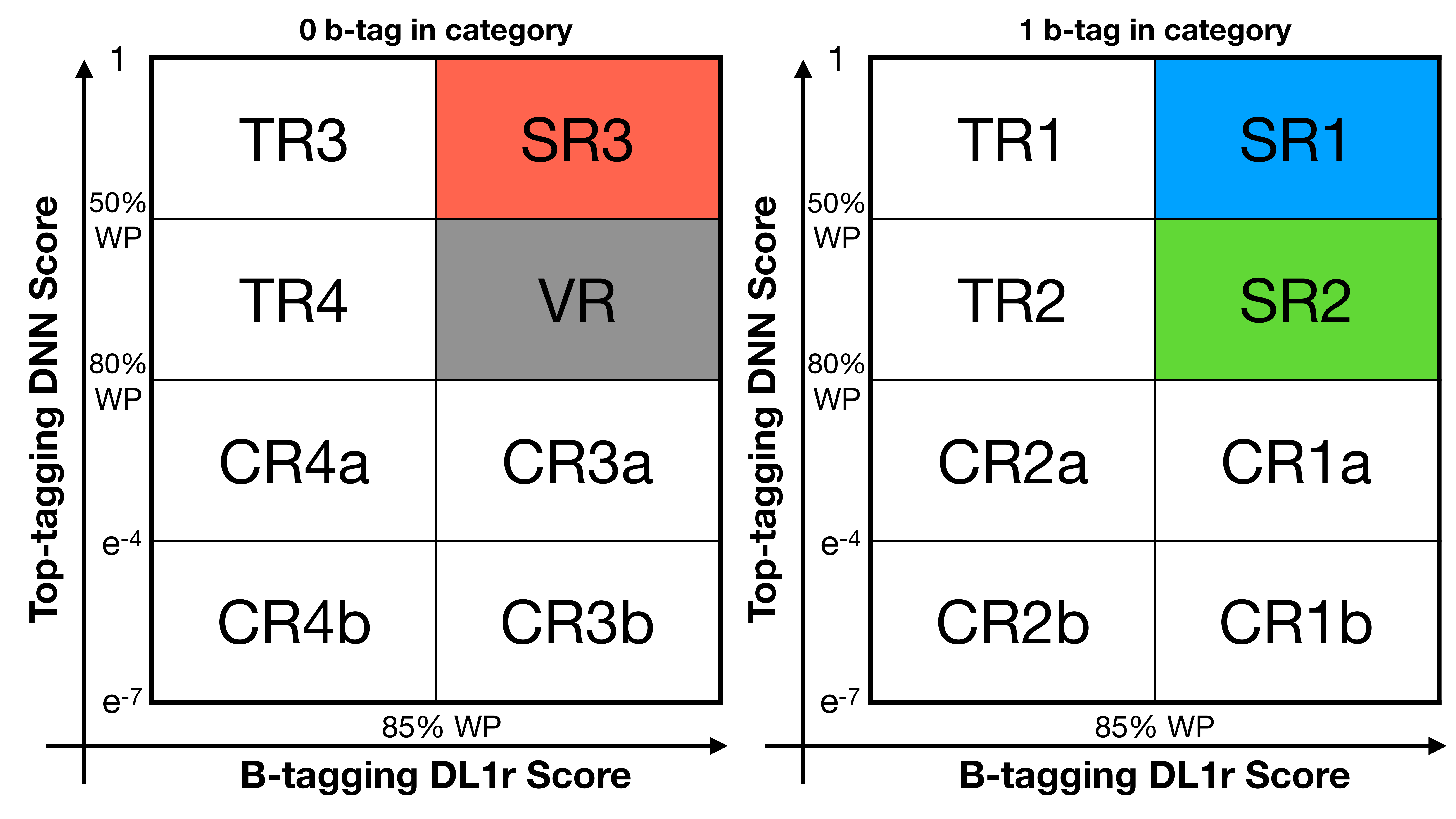}
	\caption{The Signal Regions (SR1 to SR3), the Validation Region (VR), the Template Regions (TR1 to TR4), and the Control Regions (upper: CR1a to 4a and lower: CR1b to CR4b) are shown with the cuts and selections. Taken from Ref.~\cite{ATLAS:2021drn}.}
  \label{ABCD}
\end{figure}

The dominant background to this analysis is the QCD production of multi-jet consisting of mainly light parton jets. A data-driven estimation for this background is adopted under the assumption that the $b$-tagging probability for $b$-candidate jets is unchanged for either top-proxy or top-candidate jets. The tables in Figure~\ref{ABCD} show the regions defined by the top-tagging and $b$-tagging configuration, allowing the QCD multi-jet events in each of the Signal Regions (SR1 to SR3) to be estimated using data events from the neighboring Template Regions (TR1 to TR3) and the Control Regions (CR1a to 4a) below. These Control Regions require a top-tagging score $>$ $e^{-4}$ for the top-proxy jets such that the parton flavor compositions approach the Signal and Template Regions. The two tables differ by the number of $b$-tagged small-$R$ jets inside the top-candidate (top-proxy) jets: none for the \textbf{0 b-tag in category} and more than one for the \textbf{1 b-tag in category}. This $b$-tag requirement increases the contribution from heavy-flavor, impacting the $b$-tagging rates of the $b$-candidate jet.

For the Template Regions (TRs), the only significant background other than that from QCD multi-jets is the background from $t\bar{t}$ production. The $t\bar{t}$ events estimated by event generators have to be subtracted from the TRs as the Control Regions cannot account for them. All other backgrounds are included in the \textbf{data-driven background}, dominated by the QCD multi-jet. The DNN cut of $e^{-7}$ for the lower Control Regions -- CR1b to CR4b -- corresponds to parton flavor variations between top-proxy and top-candidate jets observed in multi-jet simulation. The double ratios between the two rows of Control Regions are calculated in data to estimate the systematic uncertainties of the data-driven estimate. The tighter 50\% WP of top-tagging is applied to top-candidate jets, leaving a signal-sensitive region SR1 and a signal depleted Validation Region for cross-checking the data-driven method before unblinding data in the SRs. 

\section{Post-fit invariant mass distribution}
\begin{figure}[tph]
  \centering
  \includegraphics[width=0.5\textwidth]{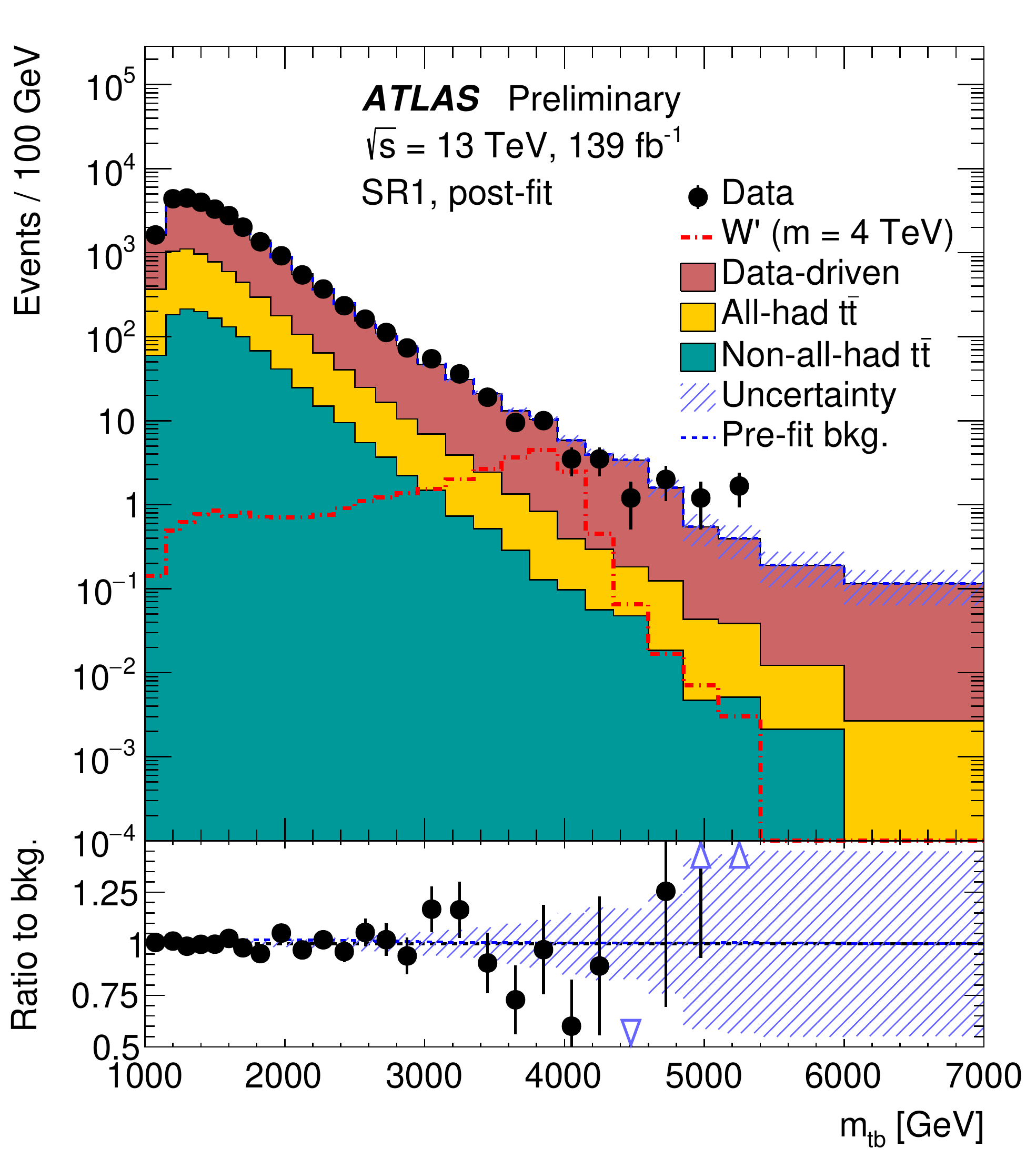}
	\caption{Post-fit invariant mass distribution in region SR1. The blue hatched band shows the systematic plus statistical uncertainty of the total background post-fit. The pre-fit total background is overlayed by blue dashed lines. The $\Wp_{R}$ shown in the red dashed histograms assume the expected cross-section. Taken from Ref.~\cite{ATLAS:2021drn}.}
  \label{SR1}
\end{figure}

The data-driven background and the MC-estimated $t\bar{t}$ background are fit to the $m_{tb}$ distribution in data with a profile-likelihood function to constrain systematic uncertainties. The post-fit distribution is compared with data in Figure~\ref{SR1} for the region SR1. The data-driven background (pale red) is plotted above the all-hadronic $t\bar{t}$ (green) and then the semileptonic plus dileptonic $t\bar{t}$ (yellow). Their sum is consistent with data adding the uncertainty bands, meaning that no significant presence of the $\Wp$ signal is observed. The overlayed right-handed $\Wp$ signal is not included in this background-only fit.

\section{Exclusion upper limit}
\begin{figure}[tph]
  \centering
  \includegraphics[width=0.6\textwidth]{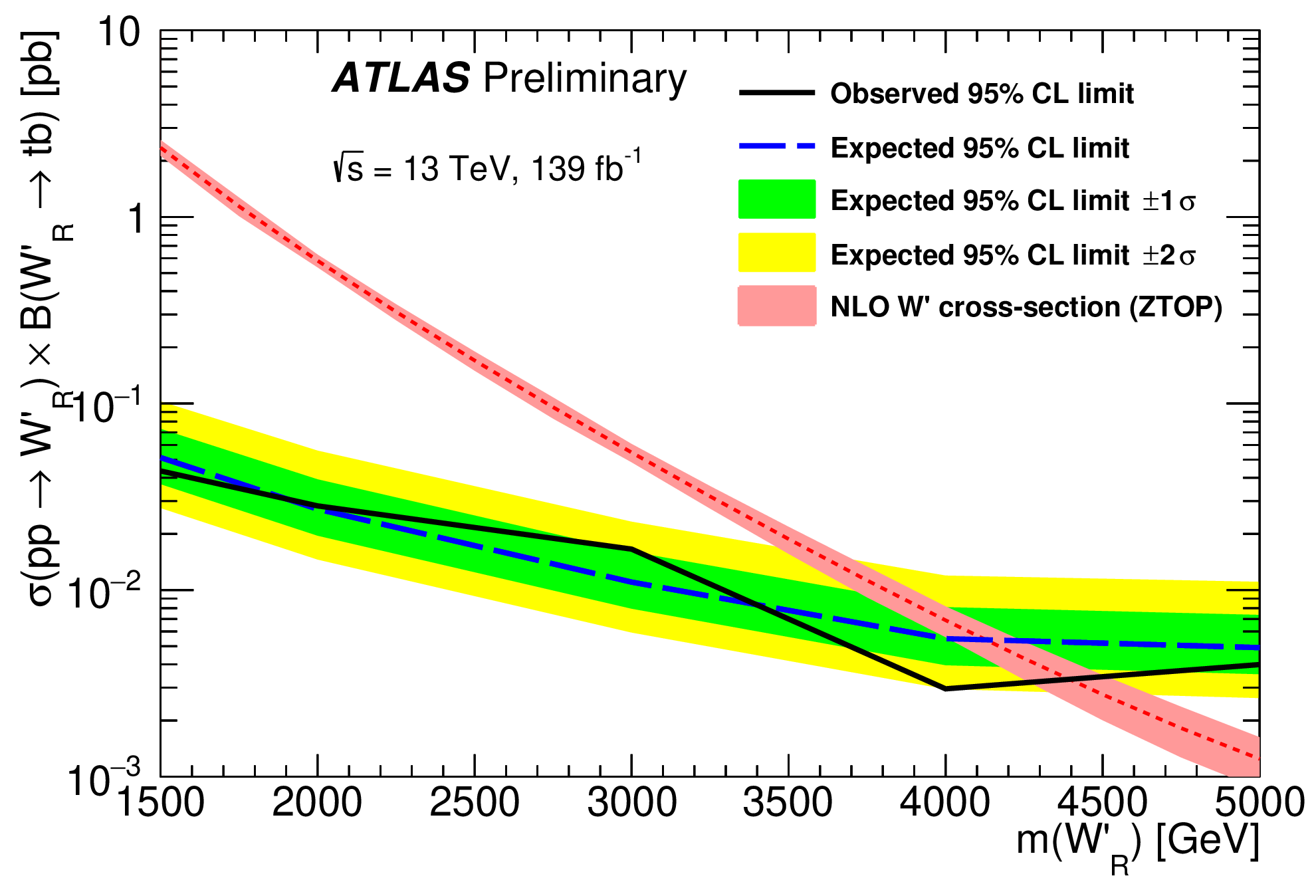}
  \caption{The exclusion upper limit on the $\Wp_R$ production times the $tb$ decay branching ratio at the 95\% confidence level as a function of the $\Wp_R$ mass. The red band includes the theory uncertainties from Parton Distribution Functions, the strong coupling constant, renormalization and factorization scale, and the top quark mass. Taken from Ref.~\cite{ATLAS:2021drn}.}
  \label{limit}
\end{figure}

We compute the upper exclusion limit for several $\Wp$ masses from 1.5 to 5 TeV at the 95\% Confidence Level, as depicted in Figure~\ref{limit}. The black line is the observed limit calculated with data; the blue dashed line is the expected limit obtained by treating the pre-fit background as expected data. The green (yellow) band corresponds to the one-sigma (two-sigma) uncertainties for the expected limit. The red line follows the theoretical cross-section for a right-handed $\Wp$ with SM-like electroweak coupling, computed by the ZTOP framework~\cite{Sullivan:2002jt,Duffty:2012rf}. As the observed limit excludes cross-sections above the curve, the $\Wp_{R}$ signal is excluded at the 95\% Confidence Level up to 4.4 TeV. For more details about this analysis, please refer to the conference note~\cite{ATLAS:2021drn}.

\bibliography{eprint}{}
\bibliographystyle{unsrt}
 
\end{document}

%% file: econfmacros.tex
\def\beq{\begin{equation}}
\def\eeq#1{\label{#1}\end{equation}}
\def\eeqn{\end{equation}}

\def\beqa{\begin{eqnarray}}
\def\eeqa#1{\label{#1}\end{eqnarray}}
\def\eeqan{\end{eqnarray}}

\let\bar=\overbar

\def\Dslash{\not{\hbox{\kern-4pt $D$}}}
\def\dslash{\not{\hbox{\kern-2pt $\del$}}}

\def\msb{{\bar{\ssstyle M \kern -1pt S}}}

